\documentclass[%
twocolumn,
showpacs,
bibnotes,
 amsmath,amssymb,
 aps,
prb,
]{revtex4-1}

\usepackage{color}

\usepackage{graphicx}
\usepackage{dcolumn}
\usepackage{bm}

\begin{document}


\title{Interplay between spin-orbit coupling and crystal-field effect in topological insulators}

\author{Hyungjun Lee}
 \affiliation{Institute of Theoretical Physics, Ecole
   Polytechnique F\'{e}d\'{e}rale de Lausanne (EPFL), CH-1015 Lausanne,
   Switzerland}
\author{Oleg V. Yazyev}%
\affiliation{%
Institute of Theoretical Physics, Ecole
   Polytechnique F\'{e}d\'{e}rale de Lausanne (EPFL), CH-1015 Lausanne,
   Switzerland}%

\date{\today}

\begin{abstract}
Band inversion, one of the key signatures of time-reversal invariant
topological insulators (TIs), arises mostly due to
the spin-orbit (SO) coupling. 
Here, based on \textit{ab initio} density-functional
calculations, we report a theoretical investigation of the SO-driven
band inversion in isostructural bismuth and antimony chalcogenide TIs from the viewpoint of
its interplay with the
crystal-field effect. We calculate the SO-induced
energy shift of states in the top valence and bottom conduction manifolds
and reproduce this behavior using a simple one-atom model adjusted
to incorporate the crystal-field effect.
The crystal-field splitting is shown to compete with the SO coupling, that is, stronger crystal-field splitting leads to weaker SO band shift.
We further show how both these effects can be controlled by changing the chemical composition, whereas the crystal-field splitting can be tuned by means of uniaxial strain.
These results provide a practical guidance to the rational design of novel TIs as well as to controlling the properties of existing materials.
\end{abstract}

\maketitle


\section{\label{sec:1}Introduction}

Topology, originally one of the branches of mathematics, has been
intimately linked to a variety of novel quantum phenomena in condensed matter
physics\cite{Monastyrsky2006,Altland2010}. One representative example is the integer quantum Hall effect
(IQHE), which was observed in the
two-dimensional (2D) electron gas in a strong magnetic
field\cite{Klitzing1980} and subsequently explained by
using the topological structures of electron wave
functions in momentum space\cite{Thouless1982}.
More recently, topological notions have
been invoked to clarify an exotic new quantum phase in
time-reversal (TR) invariant band insulators, dubbed
topological insulators
(TIs)\cite{Hasan2010,Qi2011,Moore2010,Bernevig2013,Ando2013}. Unlike the
  IQHE, TIs preserve the TR symmetry and are
topologically classified via the $\mathbb{Z}_2$ invariant\cite{Kane2005Z2,Moore2007,Fu2007PRL,Roy2009}.
As guaranteed by bulk-boundary correspondence,
this bulk topological nature
manifests itself in the robust spin-helical metallic boundary modes which are protected
by the TR symmetry\cite{Hasan2010,Qi2011,Moore2010,Bernevig2013,Ando2013}.

Nontrivial topology of TIs is accompanied by band
inversion and TIs discovered to date theoretically or experimentally can be classified into $s$-$p$, $p$-$p$, and $d$-$f$
types according to the orbital characters of states participating in
band inversion\cite{Zhu2012,Zhang2013}. The examples of TIs with $s$-$p$ type band inversion include HgTe quantum wells\cite{Bernevig2006}, the first experimentally realized 2D TI (also called quantum spin
Hall insulator)\cite{Konig02112007}, as well as ternary semiconductors
such as Heusler\cite{Chadov2010,Lin2010,Xiao2010} and chalcopyrite compounds\cite{Feng2011}.
As an example, HgTe features the inverted band order, that is, the
$s$-like $\Gamma_6$ state lies below the $p$-like $\Gamma_8$ states at the $\Gamma$ point\cite{Bernevig2006}.
The $p$-$p$ type band inversion can be found in various compounds
containing heavy $p$-block elements, notably Bi with the large atomic
spin-orbit (SO) coupling constant of 1.25~eV for the valence $6p$ states \cite{Wittel1974}. These materials include Bi$_{1-x}$Sb$_x$ alloy\cite{Fu2007PRB}, the
first experimentally identified three-dimensional (3D)
TI\cite{Hsieh2008}, the second confirmed 3D TIs of binary Bi chalcogenides (Bi$_2$Se$_3$ and Bi$_2$Te$_3$)\cite{Zhang2009,Xia2009,Chen2009}, and their variants\cite{Yan2012} such as TlBiSe$_2$\cite{Lin2010PRL,Sato2010}, PbBi$_2$Te$_4$\cite{Menshchikova2011,Souma2012}, and Bi$_2$Te$_2$Se classes\cite{Ren2010,Taskin2011}.
Among them, Bi$_2$Se$_3$ and Bi$_2$Te$_3$ along with Sb$_2$Te$_3$
commonly exhibit the band inversion at the $\Gamma$ point between two $p_z$-type states nearest
to the Fermi energy ($E_\text{F}$) with opposite parities, thus
leading to topologically nontrivial phase\cite{Zhang2009}.
Lastly, the $d$-$f$ type band inversion mostly appears in correlated materials such as actinide compounds\cite{Zhang2012} and mixed-valence compounds\cite{Lu2013,Deng2013}; for
instance, the band inversion in mixed-valence SmB$_6$ occurs between the 4$f$
and 5$d$ states, such that one 5$d$ state moves below the 4$f$ states
at three $X$ points\cite{Lu2013}.
In the examples above, band inversion occurs mostly due to the SO coupling. For this reason, all discovered TIs have the high-$Z$
elements with the strong atomic SO coupling as their constituents. Materials involving heavy elements thus constitute natural candidates to host
topologically nontrivial phases.

Typically, the SO interaction induces splitting in band structures in
various ways by
coupling the electron's spin and orbital degrees
of freedom\cite{Winkler2003}. The SO-induced splitting in solids can be
broadly divided into the
following two classes according to the presence or absence of
inversion symmetry: (1) in inversion-symmetric materials such as
diamond-type semiconductors, the SO interaction produces the band
splittings, giving rise to, for example, the SO gap and the heavy-hole-light-hole splitting near the valence-band
edge\cite{Winkler2003}; (2) in inversion-asymmetric materials such as
zinc-blende-type semiconductors and metallic surfaces, it results in
the additional spin splittings, for example, the Dresselhaus\cite{Dresselhaus1955} and
Rashba-Bychkov\cite{Bychkov1984} effects.
In its extreme manifestation, the SO interaction
leads to the nontrivial electronic topology in TIs through band inversion. In this sense, the SO
coupling has been considered as the essential ingredient to realize the
topological order in TIs by taking the role of magnetic field in
IQHE\cite{Hasan2010,Qi2011,Moore2010,Bernevig2013}.
Going one step further, as one of the possible
routes to enhance bulk band gap
in TIs\cite{Zhou07102014,Zhou2014}, there has been recently considerable interest in the interplay of the SO coupling with other
effects such as the electron-electron interaction\cite{Pesin2010,Zhang2012}.

In the present work, we investigate the interplay between the SO
coupling and the crystal-field effect in TIs by using \textit{ab initio} density-functional-theory
calculations. 
For this purpose, we consider the prototypical 3D TIs, Bi and Sb chalcogenides, all of which are isostructural
layered compounds, thus being
the ideal playground for exploring the interplay between the SO
coupling and the crystal-field effect. 
The SO-induced shifts of the valence and conduction bands are responsible for band inversion in the above-mentioned materials, which eventually results in the emergence of TI phase. 
We introduce a simple model based
on one-atom system incorporating the crystal-field effect.
This model clearly demonstrates that the magnitude of the SO-driven shift of the valence and conduction band states
is intimately related to the crystal-field splitting of these states.
Moreover, we show how both the SO coupling and the crystal-field effect can be controlled by changing the chemical composition, whereas the latter effect can be tuned separately by means of uniaxial strain. 
The results of our work provide a practical guidance to the rational design of novel TIs as well as to controlling the properties of existing materials. 

The rest of the manuscript is organized as follows: In Sec.~\ref{sec:2}, we describe our density-functional-theory computational methodology.
Sec.~\ref{sec:3} discusses the main results of our work. Namely, it
introduces the simple atomic picture of the interplay between the SO
coupling and the crystal-field effect, then discusses the first-principles calculations carried out on TIs of different chemical composition as well as the effects of uniaxial pressure. Sec.~\ref{sec:4} concludes our paper.

\section{\label{sec:2}Computational Methodology}

Our present calculations are based on first-principles
density-functional-theory (DFT) methods\cite{Hohenberg1964,Kohn1965}
as implemented in the Quantum ESPRESSO (QE)
package\cite{Giannozzi2009}. We employ the
Perdew-Burke-Ernzerhof-type generalized gradient approximation for the
exchange-correlation energy\cite{Perdew1996} and the
norm-conserving pseudopotentials (PPs)\cite{Hamann1979} generated according
to the scheme of Troullier and
Martins\cite{Troullier1991} as implemented in the APE code\cite{Oliveira2008}.
Wave functions are expanded using plane waves with kinetic-energy 
cutoff of 45~Ryd for all compounds considered. The Brillouin zone (BZ) is sampled with 
8$\times$8$\times$8 Monkhorst-Pack meshes of 
special $\vec{k}$ points\cite{Monkhorst1976}. 
All used parameters were carefully checked to ensure the convergence
of total energies to 1~mRyd. 

To treat the SO coupling, two methods are used separately, both of which rely
on the fully-relativistic PPs\cite{Kleinman1980}:
(1) for electronic structure calculations, one-shot perturbative treatment of the SO coupling based on the Hybertsen and Louie
(HL) approach\cite{Hybertsen1986} with the perturbed SO Hamiltonian constructed using the wave functions
from the scalar-relativistic calculations as unperturbed wave
functions; (2) for atomic structure optimization calculations, full and self-consistent incorporation of the SO coupling with two-component spinor wave functions as
implemented in QE\cite{Corso2005}.
The SO-driven band shift was calculated perturbatively via the HL approach as it is ambiguous how to determine the reference between the energies without and with the SO coupling. Thus, this perturbative approach constitutes a valuable tool for addressing 
the role of SO coupling separately from other effects. 
In order to assess the error due to the
non-self-consistency in the HL formalism, we calculated electronic band
structures from both methods and found that within the energy
window of 3~eV centered around the Fermi level their differences are less than 0.1~eV for Bi
chalcogenides while they are on the order of meV for Sb chalcogenides.
Both the fully self-consistent approach \cite{Yazyev2010} and the HL method \cite{Kioupakis2010,Yazyev2012} have been successfully applied to studying the bismuth chalcogenide TIs.
For structural optimizations, we adopted the full treatment of the SO
coupling.
We used the experimentally determined crystal 
structures for all examined materials\cite{Nakajima1963,Anderson1974}
with the exception of Sb$_2$Se$_3$ for which we optimized the lattice constants and
internal coordinates due to the lack of the crystallographic data for
the rhombohedral phase.
In order to address the effects of uniaxial pressure, we varied the lattice constant along
the $c$ axis in the hexagonal unit cell and then fully relaxed the internal coordinates until the residual forces on all ions are less than 1~meV/\AA.

\section{\label{sec:3}Results}

\subsection{\label{sec:3a}Atomic picture}

We start by considering the SO-induced energy splitting in the
simplest configuration, a one-atom
system with valence $p$ orbitals. It basically traces back to the
fine-structure levels of the hydrogenic atom\cite{Condon1935}.
We choose a basis set of $\{|p_x,\uparrow\rangle,\,|p_x,\downarrow\rangle,\,|p_y,\uparrow\rangle,\,|p_y,\downarrow\rangle,\,|p_z,\uparrow\rangle,\,|p_z,\downarrow\rangle\}$,
where $p_i$'s $(i=x,y,z)$ are the radial part of valence wave
functions times the corresponding real spherical harmonics and
$\uparrow\,(\downarrow)$ is the eigenstate of the
$z$ component of the spin angular momentum operator.
The total Hamiltonian with the SO interaction taken into account is
then given
by
\begin{equation}
  \label{eq:1}
  H=\frac{\zeta}{2}
  \begin{pmatrix}
    0 & 0 & -i & 0 & 0 & -1 \\
    0 & 0 & 0 & i & 1 & 0 \\
    i & 0 & 0 & 0 & 0 & i \\
    0 & -i & 0 & 0 & i & 0 \\
    0 & 1 & 0 & -i & 0 & 0 \\
    -1 & 0 & -i & 0 & 0 & 0
  \end{pmatrix}\,,
\end{equation}
with $\zeta$ the atomic SO strength\footnote{%
The atomic SO strength, $\zeta$, of an atomic orbital
  is
  \begin{equation*}
    \zeta=\int\frac{\hbar^2}{2m_\mathrm{e}^2c^2}\frac{1}{r}\frac{dV(r)}{dr}[R(r)]^2r^2\,dr\,,
  \end{equation*}
  where $m_\mathrm{e}$ is an electron mass, $V(r)$ is a potential, and $R(r)$ is the radial part of the atomic orbital.
}.
Here, we assume that before the SO interaction is introduced, all $p$-type orbitals have the same energy due to the
central potential of nucleus (i.e., spherical
symmetry) and the absence of spin polarization. The latter is not general and, typically, is not the case for isolated atoms with unpaired electrons\cite{Oliveira2013}. However, this assumption is justified for the considered systems that are non-magnetic due to the combined effect of time-reversal and inversion symmetries similar to the case of Bi dimer\cite{Oliveira2013}.
This energy is set to 0 for simplicity.
After diagonalizing this Hamiltonian, we obtain the following
eigenenergies: $E_{j,m_j}=-\zeta$ (doublet) for $j=1/2$ with $m_j=\pm
1/2$ and $\zeta/2$ (quadruplet) for $j=3/2$ with $m_j=\pm 1/2$ and
$\pm 3/2$, where $j$ is the total angular momentum and $m_j$ is its
$z$ projection. For future reference, we list here the SO
coupling parameters from experimental measurements\cite{Wittel1974} for the
elements relevant to our study: $\zeta(\text{Bi})=1.25\,\text{eV}$, $\zeta(\text{Te})=0.49\,\text{eV}$,
$\zeta(\text{Sb})=0.40\,\text{eV}$, and $\zeta(\text{Se})=0.22\,\text{eV}$.

\begin{figure}
  \centering
  \includegraphics[width=8.6cm]{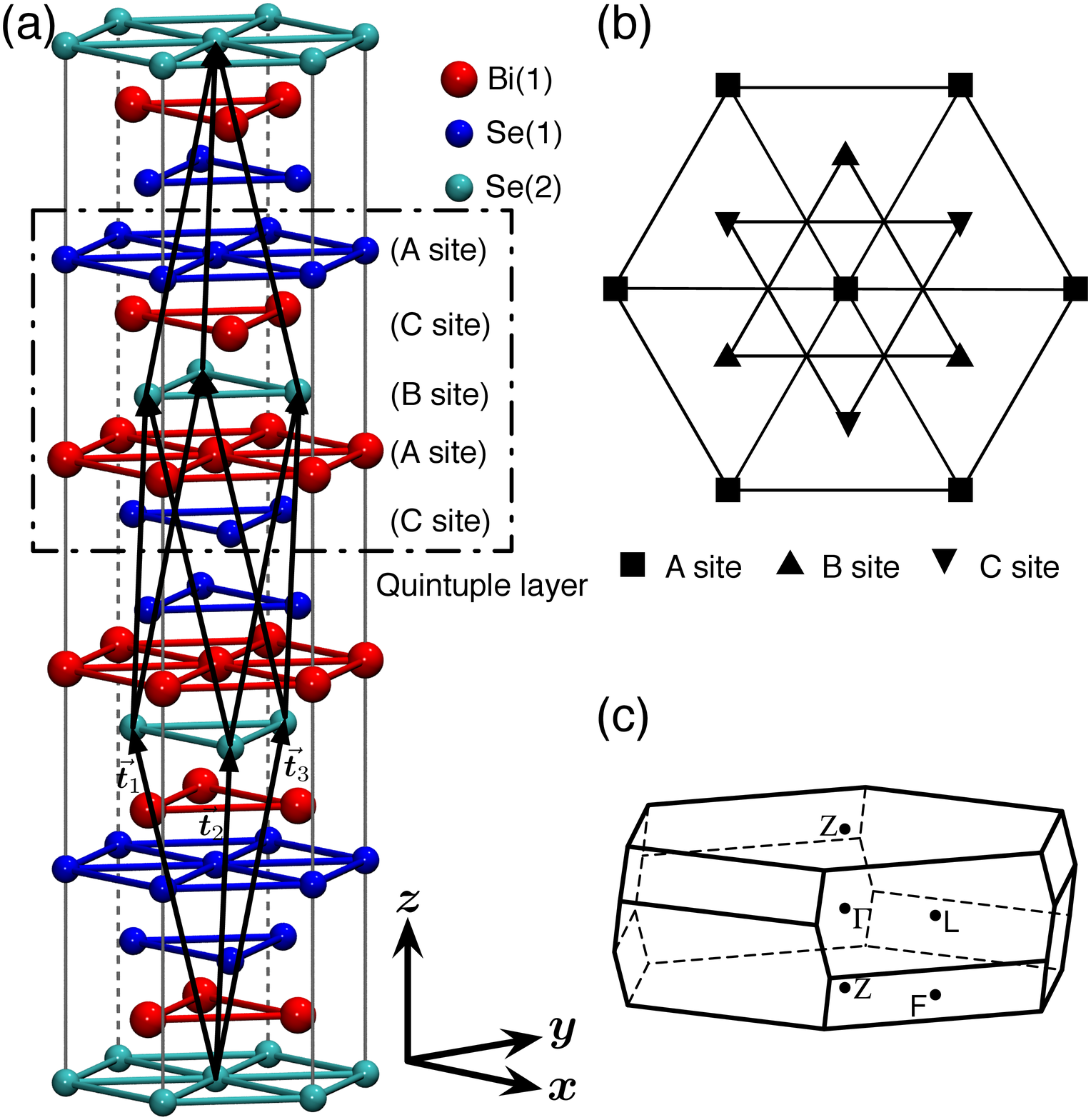}
  \caption{(Color online) 
    (a) Ball-and-stick representation of the bulk crystal structure of Bi$_2$Se$_3$ in
    which Bi atoms are indicated by red-colored balls and two
    inequivalent Se atoms by blue- (Se(1)) and cyan-colored (Se(2)) ones.
    A quintuple layer is marked by the dot-dashed rectangle with
    constituent atoms labeled according to the stacking sequence. Primitive lattice vectors
    $\vec{t}_i$ ($i=1,2,3$) are denoted with thick solid arrows and our chosen coordinate axes are also shown.
    (b) Top view along the (111) (trigonal)
    direction. Different triangular lattices are represented by rectangles, lower-,
    and upper-triangles.
    (c) First Brillouin zone (BZ) of Bi$_2$Se$_3$ bulk. Filled dots locate
    the high-symmetry $\vec{k}$ points that correspond to time-reversal invariant momenta.
  }
  \label{fig:Fig1}
\end{figure}

Next, we examine the case with the lower symmetry in which the
energies of $p$ orbitals without the SO coupling are differentiated from each other. This
applies to the atoms in Bi and Sb chalcogenides crystallizing in the rhombohedral structure with
the space group $R\bar{3}m$ ($\#166$)\cite{Nakajima1963,Anderson1974}.
With Bi$_2$Se$_3$ as an example, we show its bulk crystal structure in
Fig.~\ref{fig:Fig1}(a), the unit cell of which has five atoms, two Bi and
three Se atoms with the Se atom labeled Se(2) as the inversion center.
Of particular interest is its layered structure with a triangular lattice within each layer
[Fig.~\ref{fig:Fig1}(b)] and a quintuple layer (QL) as the
building block which is stacked along the trigonal axis with a three-fold rotational symmetry\cite{Nakajima1963,Zhang2009}.
We choose coordinate axes as follows: the $x$ axis is taken along
the binary axis with the two-fold rotational symmetry, the $y$ axis is
taken along the bisectrix axis, which is the intersection line of the
reflection plane and atomic-layer plane, and the $z$ axis is along the
trigonal axis perpendicular to the atomic layer. 

Due to the layered crystal structure and the polarity of covalent
bonds in these materials, the $p_x$ and $p_y$ orbitals are placed in
different electrostatic environments compared with the $p_z$
orbital; both the positive
($E_{p_x}=E_{p_y}>E_{p_z}$) and negative ($E_{p_x}=E_{p_y}<E_{p_z}$) crystal-field
splittings are realized\cite{Zhang2009,Liu2010}. 
The positive crystal-field splittings take place in the conduction-band states formed by the cation elements (Sb and Bi), while the
negative crystal-field splittings of the valence-band states are shared by the 
anion elements (Se or Te). This behavior is common for all investigated binary materials as well as other Bi-based compounds such as BiTeI\cite{Bahramy2011,Bahramy2012}.
Below, we examine the cases of both positive and negative crystal-field splittings.

In order to incorporate the effects of positive
crystal-field splitting ($E_\text{CF} > 0$) typical for cation (Sb and Bi) sites, the total Hamiltonian should be modified in the following way:
\begin{equation}
  \label{eq:2}
  H=
  \begin{pmatrix}
    E_{xy} & 0 & -i\zeta/2 & 0 & 0 & -\zeta/2 \\
    0 & E_{xy} & 0 & i\zeta/2 & \zeta/2 & 0 \\
    i\zeta/2 & 0 & E_{xy} & 0 & 0 & i\zeta/2 \\
    0 & -i\zeta/2 & 0 & E_{xy} & i\zeta/2 & 0 \\
    0 & \zeta/2 & 0 & -i\zeta/2 & 0 & 0 \\
    -\zeta/2 & 0 & -i\zeta/2 & 0 & 0 & 0
  \end{pmatrix}\,,
\end{equation}
where $E_{xy}$ represents the energy of the $p_x$
($p_y$) orbitals without the SO coupling and that of $p_z$ orbitals without the SO coupling is fixed to 0 for
convenience, thereby equating the crystal-field splitting $E_\text{CF}$ to $E_{xy}$. By diagonalizing this
Hamiltonian matrix, we can obtain
the SO-induced energy splitting as a function of $E_\text{CF}$ (all doublets):
\begin{equation}
  \label{eq:3}
  E_{j,m_j}=\\
  \left\{
\begin{array}{>{\displaystyle}l}
  \frac{1}{4}\left(-\zeta+2E_\text{CF}-\sqrt{9\zeta^2-4\zeta
      E_\text{CF}+4E_\text{CF}^2}\right)\\
  \hfill\mbox{for $j=1/2$ and $|m_j|=1/2$}\\
  \frac{1}{4}\left(-\zeta+2E_\text{CF}+\sqrt{9\zeta^2-4\zeta
      E_\text{CF}+4E_\text{CF}^2}\right)\\
  \hfill\mbox{for $j=3/2$ and $|m_j|=1/2$}\\
  \frac{1}{2}\left(\zeta+2E_\text{CF}\right)
  \hfill\mbox{for $j=3/2$ and $|m_j|=3/2$}
\end{array}
\right.\,.
\end{equation}
Note that both $j$ and $m_j$ are not good quantum numbers for this Hamiltonian with
non-zero crystal-field splitting, but we
use the effective $j$ and $m_j$ for classification purposes.
We also note that in the limiting case, $E_\text{CF}\rightarrow\infty$,
the energy of $j=1/2$ states shifts downward due to the SO coupling by $\zeta/4$,
i.e., one fourth of its maximum shift at zero crystal-field splitting.

The negative crystal-field splitting scenario ($E_\text{CF} < 0$) typical for the anion sites (Se and Te) is incorporated phenomenologically into the
 one-atom model by changing the sign of
$\zeta$ in Eq.~(\ref{eq:2}). It suffices for our
purposes because we are interested mainly in the shift of $p_z$-like
($j=1/2$) states closest to $E_\text{F}$ rather than the details in entire $p$ manifold. The
resulting eigenenergies are as follows:
\begin{equation}
  \label{eq:3a}
  E_{j,m_j}=\\
  \left\{
\begin{array}{>{\displaystyle}l}
  \frac{1}{4}\left(-\zeta+2E_\text{CF}+\sqrt{9\zeta^2-4\zeta
      E_\text{CF}+4E_\text{CF}^2}\right)\\
  \hfill\mbox{for $j=1/2$ and $|m_j|=1/2$}\\
  \frac{1}{4}\left(-\zeta+2E_\text{CF}-\sqrt{9\zeta^2-4\zeta
      E_\text{CF}+4E_\text{CF}^2}\right)\\
  \hfill\mbox{for $j=3/2$ and $|m_j|=1/2$}\\
  \frac{1}{2}\left(\zeta+2E_\text{CF}\right)
  \hfill\mbox{for $j=3/2$ and $|m_j|=3/2$}
\end{array}
\right.\,.
\end{equation}

\begin{figure}
  \centering
  \includegraphics[width=8.6cm]{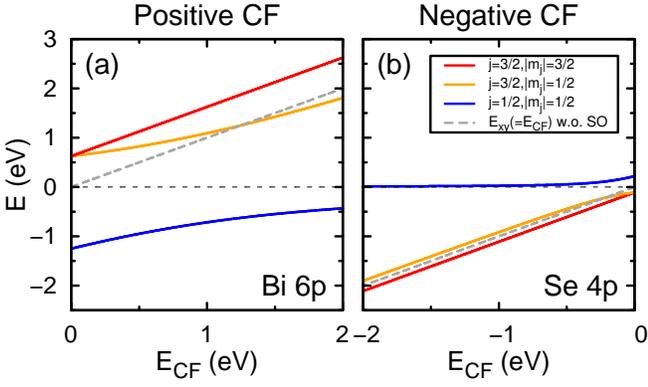}
  \caption{(Color online) Atomic energy levels as a function of 
crystal-field splitting $E_\text{CF}$ for (a) Bi $6p$ states and 
(b) Se $4p$ states at experimentally determined SO parameter values. As described in the main text,
    the positive and negative values of the crystal-field splitting 
are considered for
    Bi and Se atoms, respectively.
 All energy levels are
    labeled by the effective total angular momentum ($j$) and its $z$
    component ($m_j$).
    Gray-dashed lines represent the energies of $p_x$ ($p_y$) orbitals ($E_\text{xy}$)
    with those of $p_z$ ones fixed to 0 in absence of the SO coupling.
    Thin dashed lines indicate zero energy.
  }
  \label{fig:Fig2}
\end{figure} 

Figures \ref{fig:Fig2}(a) and \ref{fig:Fig2}(b) show the 
eigenvalues of Hamiltonian (\ref{eq:2}) as a function of  
$E_\text{CF}$ for the experimentally 
determined values\cite{Wittel1974} of SO parameter $\zeta$
of Bi $6p$ and Se $4p$ atomic states, respectively. In these figures, we focus only on positive values of $E_\text{CF}$ for the Bi $6p$ states and on negative values of $E_\text{CF}$ for the Se $4p$ states.
From these figures, one can clearly see that the SO-driven splittings
strongly depend on $E_\text{CF}$. For Bi atom [Fig.~\ref{fig:Fig2}(a)], the SO coupling first induces the
splitting between $j=1/2$ doublet and $j=3/2$ quadruplet when
$E_\text{CF}$ is zero, then quadruplet splits further into two
doublets, each with $|m_j|=1/2$ and $3/2$ as $E_\text{CF}$ increases.
More precisely, at zero crystal-field splitting the $j=1/2$ doublet is shifted downward
by its maximum magnitude of 1.25~eV, then this magnitude decreases
as $E_\text{CF}$ increases, reaching
about 0.43~eV when $E_\text{CF}$ becomes 2~eV.
In comparison
to Bi atom with the positive crystal-field splitting, there are two distinct
features in Se atom with the negative crystal-field splitting [Fig.~\ref{fig:Fig2}(b)]: one is the reversal of the order of energy
levels to arrange them from the lower energy in the following sequence, $E_{j=3/2,|m_j|=3/2}$,
$E_{j=3/2,|m_j|=1/2}$, and $E_{j=1/2,|m_j|=1/2}$; the other is the opposite sign of the
shift of the states. As compared to Bi atom, the $j=1/2$ doublet of Se
atom is shifted upward by its maximum magnitude of 0.22~eV at zero
crystal-field splitting, whereas the $j=3/2$ states are shifted downward.
These results clearly show that depending on both the magnitude
and the sign of the crystal-field
splitting, the SO-driven splitting changes correspondingly and in particular, the
maximum shift of the $j=1/2$ doublet always occurs at zero crystal-field splitting. Hence, the relative importance of the SO coupling tends to diminish with the increase of  the crystal-field splitting.
We emphasize that in the case of discussed materials the $j=1/2$
states of both cation and anion atoms are relevant to the
top valence band and
    bottom conduction band states that exchange their order upon band inversion.

\subsection{\label{sec:3b}Realistic materials}

Keeping in mind the above results for SO-induced splitting in one-atom model system, we shall now turn our attention to the SO-driven band inversion in
bismuth (Bi$_2$Se$_3$, Bi$_2$SeTe$_2$, and Bi$_2$Te$_3$) and antimony
chalcogenides (Sb$_2$Se$_3$, Sb$_2$SeTe$_2$, and Sb$_2$Te$_3$).
Except for Sb$_2$Se$_3$ with the orthorhombic phase (space group $Pbnm$) at ambient
conditions\cite{Tideswell1957}, all these materials crystallize in the
rhombohedral phase (space group $R\bar{3}m$)\cite{Nakajima1963,Anderson1974}.
For comparison purposes, we assume the rhombohedral phase also for Sb$_2$Se$_3$.

\begin{figure}
  \centering
  \includegraphics[width=8.6cm]{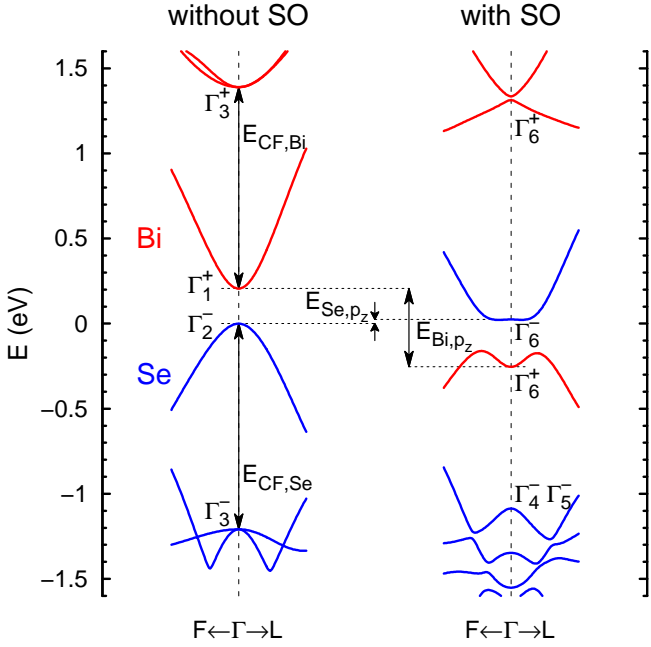}
  \caption{(Color online) 
    Band structures of bulk Bi$_2$Se$_3$ along the F-$\Gamma$-L
    path near the $\Gamma$ point without and with the SO
    interaction. The energy of the valence band maximum at $\Gamma$ without SO is set to 0.
   The energy bands are calculated along the path from
    $\Gamma$  towards both F and L points up to the distance of
    0.2~\AA$^{-1}$, and are depicted with different colors
    according to the dominant atomic character at $\Gamma$.
    The states at $\Gamma$ are labeled by the irreducible symmetry representation with the superscripts denoting the corresponding parity. The definitions of 
    relevant energies discussed in the text are indicated by the arrows.
  }
  \label{fig:Fig3}
\end{figure}

Figure~\ref{fig:Fig3} shows the electronic band structures of bulk Bi$_2$Se$_3$ near the
valence- and conduction-band edges around the $\Gamma$ point without
and with the SO coupling, 
whose main features are common also in the band structures of other members of this class.
As indicated in
this figure, the wave functions at the $\Gamma$ point can be classified
according to the irreducible representation of the space group
$R\bar{3}m$\cite{Liu2010}. 
In this layered crystal structure, as considered above for one-atom system,
the in-plane orbitals ($p_x$ and $p_y$) are differentiated from the
out-of-plane orbital ($p_z$), thus resulting in the crystal-field splitting between the $p_x$ ($p_y$)-dominated band with $\Gamma_3^\pm$
symmetry and the $p_z$-dominated one with $\Gamma_1^\pm$ or $\Gamma_2^\pm$
symmetry even before the SO coupling is considered\cite{Zhang2009,Liu2010}.
The SO coupling
further splits these band levels into three doublets, each of which
belongs to the states of $|j=1/2,|m_j|=1/2\rangle$,
$|j=3/2,|m_j|=1/2\rangle$, and $|j=3/2,|m_j|=3/2\rangle$\cite{Zhang2009,Liu2010}. Most
importantly, at the $\Gamma$ point, the SO coupling reverses the band
ordering between the top
valence ($\Gamma_2^-$) and bottom conduction ($\Gamma_1^+$) bands with opposite parities,
thereby realizing topologically nontrivial phase\cite{Zhang2009,Pertsova2014}.

\begin{figure}
  \centering
  \includegraphics[width=8.6cm]{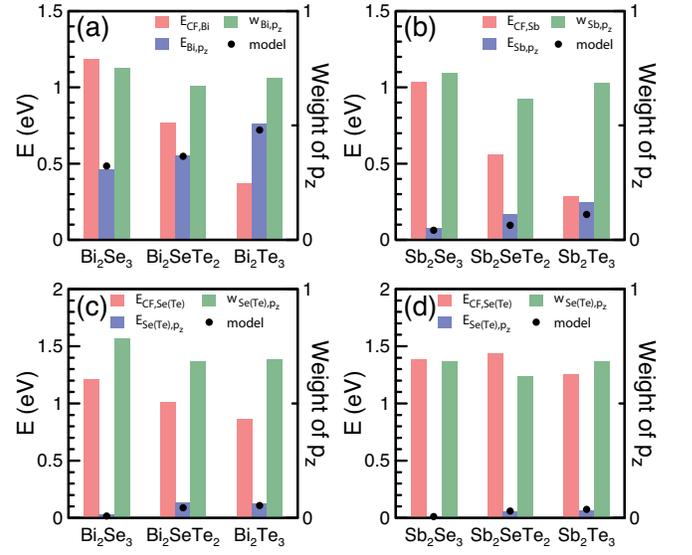}
  \caption{(Color online) 
    Bar diagrams revealing the effects of crystal field on SO-driven band
    shifts along a series of (a,c) Bi chalcogenides and (b,d) Sb chalcogenides. The depicted quantities are the
    crystal-field splittings, $E_\text{CF,Bi(Sb)}$
      or $E_\text{CF,Se(Te)}$, of the conduction- or valence-band edges, the
    weight of $p_z$ orbital, $w_{\text{Bi(Sb)},p_z}$ ($w_{\text{Se(Te)},p_z}$), of cation (anion) atom in the bottom
    conduction (top valence) band at $\Gamma$ without the SO coupling, and
    the magnitude of its SO-induced shift,
      $E_{\text{Bi(Sb)},p_z}$ and $E_{\text{Se(Te)},p_z}$, in
    presence of the SO coupling. Filled
    dots indicate the
    SO-driven shifts calculated using the one-atom model introduced in Sec.~\ref{sec:3a}.
  }
  \label{fig:Fig4}
\end{figure}

The strength of the SO coupling and the crystal-field effect in this class of materials
may be different depending on distinct combinations
of the anion and cation atomic species.
The former is inherited from 
the atomic SO coupling while the latter depends on the coordination environments of constituent atoms and their relative electronegativities. 
As explained above, the SO-driven band shifts result from the interplay between these two factors.
In order to verify this, for different
combinations of elements we obtain at the $\Gamma$ point the
crystal-field splitting, $E_\text{CF,Bi(Sb)}$ or $E_\text{CF,Se(Te)}$, in cation (Bi, Sb)- or anion (Se, Te)-derived
conduction- or valence-band edge and the weight, $w_{\text{Bi(Sb)},p_z}$ ($w_{\text{Se(Te)},p_z}$), of cation (anion)
$p_z$ orbital in bottom conduction (top valence) state before the SO
coupling is introduced; we then calculate perturbatively the shift of
these $p_z$-type levels, $E_{\text{Bi(Sb)},p_z}$ and $E_{\text{Se(Te)},p_z}$, at the $\Gamma$ point after the
SO coupling is included. These results are shown for Bi and Sb
chalcogenides in Figs.~\ref{fig:Fig4}(a,c) and \ref{fig:Fig4}(b,d), respectively.

The substitution of Te for Se in Bi chalcogenides decreases the crystal-field splitting, likely because of the smaller  electronegativity difference between Bi and Te compared to the Bi-Se pair. In Bi-derived conduction-band edge such substitution increases the shift of Bi
$p_z$-type lowest conduction level ($E_\text{Bi,$p_z$}$) when the SO coupling  is included [Fig.~\ref{fig:Fig4}(a)].
More precisely, for Bi
chalcogenides, the substitution of Te for Se decreases the crystal-field
splitting, $E_\text{CF,Bi}$ and $E_\text{CF,Se(Te)}$, from 1.18 to 0.37~eV and from 1.21 to 0.86~eV in the bottom conduction and the top valence manifolds, respectively. In turn, owing to these changes of the
crystal-field splitting, the SO-driven shifts, $E_{\text{Bi},p_z}$ and $E_{\text{Se(Te)},p_z}$, of Bi $p_z$ and Se
(Te) $p_z$ states 
increase from 0.46 to 0.76~eV and from 0.02 to 0.13~eV, respectively, both with the highest values in Bi$_2$Te$_3$.
On the other hand, during these substitutions, the weights of Bi $p_z$
and Se (Te) $p_z$ states, $w_{\text{Bi},p_z}$ and $w_{\text{Se(Te)},p_z}$, change
little in bottom
conduction and top valence states, respectively, and do not show any systematic tendency. The results for Sb
chalcogenides in Figs.~\ref{fig:Fig4}(b) and \ref{fig:Fig4}(d) show
the same qualitative trends. However, due to the factor of 3 smaller atomic SO strength in Sb, the shift of
Sb $p_z$-type level, $E_\text{Sb,$p_z$}$, in Sb chalcogenides is much smaller than $E_\text{Bi,$p_z$}$ in Bi chalcogenides.

Now, we briefly digress to recall the results obtained using the simple one-atom model. 
This model is expected to provide reasonable estimates of SO-driven splitting
 because the SO coupling is much more
prominent deep in the core region,
thus it can be safely approximated using the $\vec{k}$-independent term of
the SO coupling in the $\vec{k}\cdot\vec{p}$ approximation analogous to the atomic SO coupling form\cite{Kane1959}.
In order to verify this argument we estimate the SO-driven shifts of Bi and Sb
$p_z$-type levels nearest the Fermi level from the corresponding one-atom results, and compare these values
with those obtained above. In this estimation, the SO-driven shift
in real materials is
obtained by multiplying the SO-induced shift of the $p_z$-dominant level
($j=1/2$ doublet) from one-atom model
by the weight of $p_z$ character of the corresponding atom in bottom conduction or top valence state at the $\Gamma$ point from DFT
results for real materials without the SO coupling.
For instance, the estimated SO-driven band shift $\tilde{E}_{\text{Bi},p_z}$ of Bi
  $p_z$-like state in Bi chalcogenides is calculated by using the
  following expression: $\tilde{E}_{\text{Bi},p_z}=|E_{\text{Bi};j=1/2,|m_j|=1/2}|\times w_{\text{Bi},p_z}$.
As illustrated in Fig.~\ref{fig:Fig4}, even though we used the minimal model
based on one-atom system, the obtained results are
consistent with those calculated using DFT with the SO interaction included using the HL method. The maximum deviation of
0.08~eV is observed for Sb$_2$Te$_3$. This rather large deviation can be ascribed to the relative large contribution of Te
(31\%), with its $\zeta$ similar to that of Sb, to the bottom
conduction band without the SO coupling.

\begin{figure}
  \centering
  \includegraphics[width=7cm]{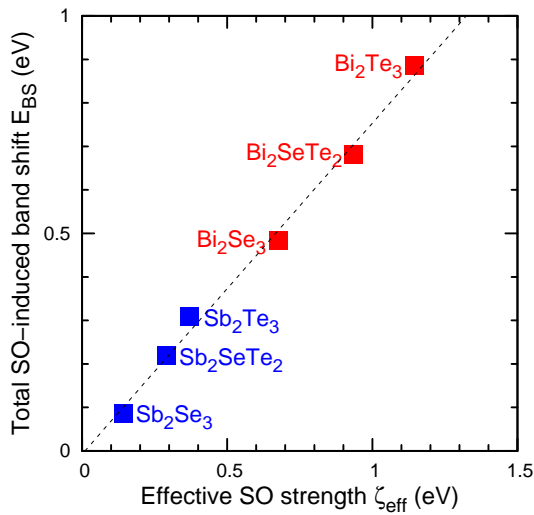}
  \caption{(Color online) 
    Relation between the effective
    spin-orbit strength $\zeta_\text{eff}$ and the
      total SO-induced band shift
    $E_\text{BS}$ in Bi and Sb chalcogenides. 
    Dashed line is a fit to the data.
  }
  \label{fig:Fig5}
\end{figure}

Relying on the existence of intimate relation between the
SO-driven splitting and the
crystal-field splitting, we can introduce an effective parameter
which can be useful for assessing the SO-driven band shift in solids much as the
atomic SO strength $\zeta$ does in atomic
systems.
Since there exists an inverse relation between the SO-induced band
shift and the crystal-field splitting relevant to this band,
we propose the following definition for the effective SO strength
\begin{equation}
  \label{eq:4}
  \zeta_\text{eff}=\sum_i \frac{\zeta_i}{1+ \frac{E_{\text{CF},i}}{\zeta_i}}\,,
\end{equation}
where $i$ refers to the atomic species index, $\zeta_i$ is
the atomic SO strength for the atom of species $i$, and
$E_{\text{CF},i}$ is the crystal-field splitting relevant to the bands
dominated by this atom
near $E_\text{F}$. Then, we calculate the total SO-induced shift of the top valence and bottom conduction band states relative to each other 
$E_\text{BS}$, e.g. for Bi$_2$Se$_3$, as 
$E_\text{BS}=E_{\text{Bi},p_z}+E_{\text{Se},p_z}$.

Figure~\ref{fig:Fig5} shows $E_\text{BS}$ as a function of
$\zeta_\text{eff}$ for Bi and Sb chalcogenides.
Nearly linear relation between these two quantities is clearly visible in
this figure.
One can also see that the atomic SO strength is not
fully utilized in solids and this degree of utilization is determined
by the relevant crystal-field splitting.
In this context, we note that our proposed effective SO strength can
be used in solids as one
of the alternatives instead of $\zeta$.

\subsection{\label{sec:3c}Effect of uniaxial strain on band inversion}

\begin{figure}
  \centering
  \includegraphics[width=8.6cm]{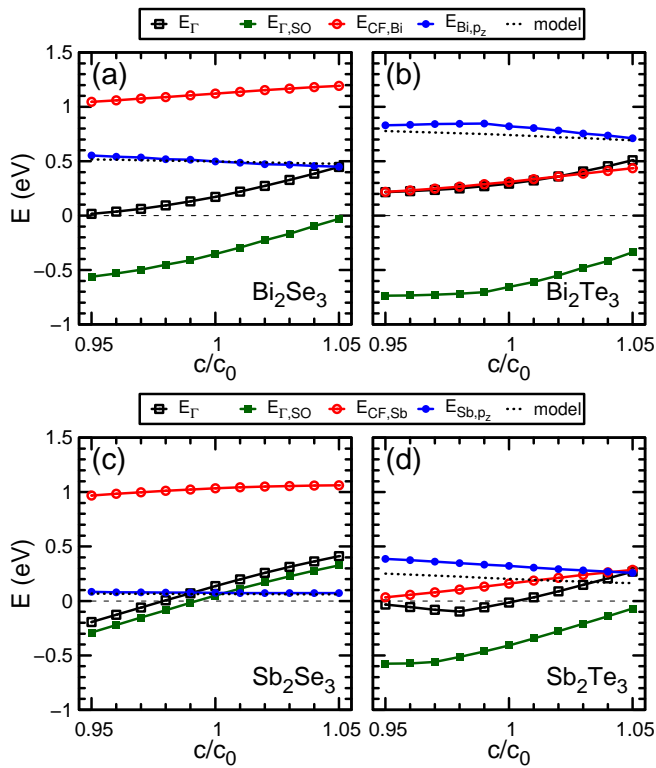}
  \caption{(Color online) 
    Effect of uniaxial strain on band inversion in (a,b) Bi chalcogenides and (c,d) Sb
    chalcogenides.
    The evolution of the band gap ($E_\Gamma$) and the crystal-field
    splitting ($E_\text{CF,Bi}$ and $E_\text{CF,Sb}$) in the bottom conduction manifold at $\Gamma$ without SO is shown as a function of uniaxial strain
    along the $z$ axis.
    We also show the evolution of the SO-driven band shift
    ($E_{\text{Bi},p_z}$ or $E_{\text{Sb},p_z}$) of Bi or
    Sb $p_z$-like state closest to $E_\text{F}$ and the SO-induced band
    gap ($E_{\Gamma,\text{SO}}$) at $\Gamma$.
    The SO-driven shift calculated using the one-atom
    model is indicated with dotted lines.
    }
  \label{fig:Fig6}
\end{figure}

Finally, we consider the effect of strain on band
inversion in the context of its intermediate role between the SO coupling and the crystal-field effect.
Compressive or tensile strain, indeed, has been considered as one of the methods to realize the
topological phase
transition through band inversion\cite{Chadov2010,Yang2010,Bahramy2012,Xi2013}; the crystal-field splitting can also be tuned by
applying pressure to materials\cite{Bahramy2012}. In this regard, we
examine, according to the applied
uniaxial strain, the
evolution of the band gap ($E_\Gamma$), the crystal-field splitting in
conduction-band edge, the weight of Bi- or Sb-$p_z$ character in Bi or
Sb $p_z$-like state nearest $E_\text{F}$ at $\Gamma$ without SO, its SO-driven band shift,
and the SO-induced band gap ($E_{\Gamma,\text{SO}}$) at $\Gamma$.
These results are shown in Fig.~\ref{fig:Fig6} for Bi chalcogenides
(Bi$_2$Se$_3$ and Bi$_2$Te$_3$) and Sb chalcogenides (Sb$_2$Se$_3$ and Sb$_2$Te$_3$).
In all cases, we apply uniaxial pressure along the (111) direction, the $z$ direction
in our chosen coordinate system. For this purpose, we first vary the $z$
component of lattice vector ($c$) in the hexagonal cell from 95~\% to 105~\% with respect to its
equilibrium value ($c_0$) and then optimize the internal atomic coordinates
until the residual forces on all ions are less than 1~meV/\AA.

As shown in Fig.~\ref{fig:Fig6}, for all considered materials strain clearly changes the
crystal-field splitting: for compressive strain, the crystal-field
splitting decreases, while for tensile strain, it increases. This
dependence can be rationalized from the following standpoint: compressive strain increases the local symmetry of coordination spheres of atoms in highly anisotropic layered materials. The change of
crystal-field splitting upon varying strain is rather small, which in
turn induces a relatively small change of SO-driven band shift. For instance, in
Bi$_2$Se$_3$ [Fig.~\ref{fig:Fig6}(a)], the crystal-field splitting in
Bi-dominant bottom conduction manifold without SO ($E_\text{CF,Bi}$) gradually increases from 1.05 to 1.19~eV as strain changes $c/c_0$
from 0.95 to 1.05. Under this rather small change in
$E_\text{CF,Bi}$, the SO-driven shift of Bi $p_z$-like state closest to
$E_\text{F}$ ($E_\text{Bi,$p_z$}$) varies inversely proportional to $E_\text{CF,Bi}$ from 0.45 to 0.55~eV.
As compared to Bi$_2$Se$_3$, the much smaller crystal-field splitting
in Bi$_2$Te$_3$ gives rise to the
larger SO-driven shift of Bi $p_z$-type level than in Bi$_2$Se$_3$ ranging from
0.71 to 0.83~eV [Fig.~\ref{fig:Fig6}(b)]. For Sb chalcogenides
[Figs.~\ref{fig:Fig6}(c) and 6(d)], 
within the whole pressure range investigated,
the smaller $\zeta$ in Sb
atom results in the smaller SO-induced band shift of Sb $p_z$-like state
closest to $E_\text{F}$ than that in the corresponding Bi chalcogenides.
We also notice that the model
based on the one-atomic picture can explain semi-quantitatively this
evolution of SO-driven shift of Bi or Sb $p_z$-type level as well,
with the maximum deviation of about 0.14~eV in Sb$_2$Te$_3$
from the result obtained via DFT plus HL methods. This maximum
deviation can also be explained by the non-negligible weight of Te in
Sb $p_z$-like state.

In contrast to small change in the crystal-field splitting, the band gap changes significantly over the considered pressure range
even before the SO coupling is included. Hence, we can see that the pressure-driven energy shift is of as much importance to the
topological quantum phase transition through band inversion as the
SO effect.
This prominent role of the pressure-induced
energy shift can be explained by the opposite shift of the energies of cation $p_z$
bonding and anion $p_z$ antibonding states near $E_\text{F}$ upon compression or expansion\cite{Liu2014}.
In fact, both the inter-QL distance and the
interlayer distance within a QL decrease as $c/c_0$ decreases,
thereby leading to, e.g. for Bi$_2$Se$_3$, the opposite shift of the Bi $p_z$-like bonding and Se $p_z$-like antibonding states irrespective of the
SO coupling\cite{Liu2014}. This reduction of layer distance has a
comparatively small influence on the SO-induced band shift since the SO coupling is
substantially strong near the core region of the atom.

As far as the topological phase transition is concerned, all investigated materials except for Sb$_2$Se$_3$ show the topologically
nontrivial phase throughout the considered pressure range. For Sb$_2$Se$_3$,
the topological phase transition takes place
at a compressive strain of $c/c_0=0.99$. This critical point is consistent with that in the previous literature\cite{Liu2011}.
As a final note, the upshift of $E_\Gamma$ below the point of $c/c_0=0.98$ in
Fig.~\ref{fig:Fig6}(d) for Sb$_2$Te$_3$ is attributed to the change of band
ordering near the valence-band edge between Te $p_x$ ($p_y$)-like and
Sb $p_z$-like states.

\section{\label{sec:4}Conclusion}

In summary, we investigated the SO-driven band inversion in TIs from
the viewpoint of its interplay with the crystal-field effect, based on
\textit{ab initio} density-functional calculations. We calculated perturbatively
the SO-induced energy shift of states near the Fermi level and
demonstrated that this shift depends on the crystal-field splitting. This behavior is
semi-quantitatively reproduced by the simple model based on the one-atom
system adjusted to incorporate the crystal-field effect. 
The crystal-field splitting is shown to compete with the SO coupling,
that is, stronger crystal-field splitting leads to weaker SO band shift.
We further demonstrated how both these effects can be controlled by changing the chemical composition, whereas the crystal-field splitting can be tuned by means of uniaxial strain.
Our results provide a helpful guideline for the discovery of novel materials realizing topologically nontrivial phases, or for tuning the properties of known topological insulators.

\begin{acknowledgments}
  We were supported by the Swiss NSF (Grant No. PP00P2\_133552) and the
  European Research Council starting grant ``TopoMat'' (Grant
  No. 306504). First-principles computations have been performed at the Swiss National Supercomputing Centre under Project Nos. s443 and s515.
\end{acknowledgments}

\bibliography{reference_v5}

\end{document}